\documentclass[aps,showpacs,preprintnumbers,amsmath,amssymb,nofootinbib]{revtex4}
\usepackage{bm}

\begin{document}

\title{Expansion of the universe from a 5D vacuum within a nonperturbative scalar field formalism}

\author{ Alfredo Raya\footnote{
E-mail address: raya@ifm.umich.mx}}
\affiliation{Instituto de F\'{\i}sica y Matem\'aticas, Universidad Michoacana de San Nicol\'as de Hidalgo,
Apartado Postal 2-82, C.P. 58040, Morelia, Michoac\'an, M\'exico.}

\author{Jos\'e Edgar Madriz Aguilar\footnote{E-mail address: jemadriz@fisica.ufpb.br}}
\affiliation{Departamento de F\'{\i}sica, Universidade Federal da Para\'{\i}ba. C.P. 5008, \\ Jo\~{a}o Pessoa, PB
58059-970 Brazil.}

\author{Mauricio Bellini\footnote{E-mail address: mbellini@mdp.edu.ar}}
\affiliation{Consejo Nacional de Investigaciones
Cient\'{\i}ficas y T\'ecnicas (CONICET)}
\affiliation{Departamento de
F\'{\i}sica, Facultad de Ciencias Exactas y Naturales, Universidad
Nacional de Mar del Plata, Funes 3350, (7600) Mar del Plata,
Argentina.}

\vskip .5cm
\begin{abstract}
We study the dynamics of the quantum scalar field responsible for inflation in different epochs of the evolution of the universe 
by using a recently introduced nonperturbative formalism from a 5D apparent vacuum.
\end{abstract}


\maketitle

\section{Introduction and basic formalism}

Multi-dimensional theories of gravity and, in particular, the
recent Braneworld cosmological scenario, in which the
universe of our perception is a submanifold embedded in a high
dimensional manifold, have generated very much interest in
mathematical and physical aspects \cite{au,aau,bu} of embedding
theories based on four dimensional general relativity (4DGR).
According to 4DGR, the model that describes the large scale
structure of the universe is a 4D manifold equipped with a
pseudo-Riemannian metric structure. The possibility of the
existence of more than 4D was taken into account not very long after the final
formulation of 4DGR by Kaluza and Klein (KK) \cite{KK}. In the
original theory, the extra dimension was considered as
topologically compact and the 4D spacetime was assumed independent of the fifth dimension. In this theory the universe was
considered as a hyper-cylinder, locally homeomorphic to ${\Re}^4
{\rm x} S^1$. In the modern version of the KK theory, these
assumptions are removed and the result is the unification of
gravitation and electromagnetism with induced sources (mass and
charge) \cite{OW,lwess,cqg}. Another recent cosmological scenario
claims that the universe of our perception is a brane of 4D
embedded in a $n$D bulk of at least 5D \cite{rs}. The main ansatz
is that matter is confined on the brane and only
gravity and electromagnetic fields can propagate on the bulk.
Mathematically, these theories are based on a basic theorem of
Riemannian geometry due to Campbell, who stated it for the first
time \cite{campbell}, and Magaard \cite{magaard}, who presented a
strict proof. This is a local embedding theorem where every
smooth analytic $n$D manifold can be locally and isometrically
embedded into some other $(n+1)$D one, which is Ricci-flat,
$R_{AB}=0$ (In our conventions, capital Latin indices $A,\ B,\ldots$ run from 0 to 4, whereas greek indices $\mu,\ \nu, \ldots $ 
from 0 to 3). In a sense, this is equivalent to the local
embedding of Einstein 4DGR, into some 5D relativity in
vacuum where matter does not exist. The idea of this approach is to
explain matter in 4D as a manifestation of pure geometry in higher
dimensions.
In a cosmological framework, scalar fields have been recognized to be responsible for the expansion of the
universe \cite{jd}, and have been proposed to
explain inflation \cite{Guth}, as well as the present accelerated
quintessential expansion \cite{we}.
The main goal of this paper is  the study of a
power-law expansion of the universe from a 5D apparent vacuum, defined
by a 5D flat metric and a purely kinetic Lagrangian for a scalar field
minimally coupled to gravity.
For this purpose, we consider the 5D metric introduced in \cite{PLB,madbe},
\begin{equation}\label{6}
dS^2 = \epsilon\left(\psi^2 dN^2 - \psi^2 e^{2N} dr^2 - d\psi^2\right),
\end{equation}
where $dr^{2}=dx^{2}+dy^{2}+dz^{2}$. The coordinates ($N$,$r$) are dimensionless, the fifth coordinate $\psi $ has spatial units  
and $\epsilon$ is a dimensionless parameter that can take the values $\epsilon = \pm 1$, accounting for the two possible 
signatures of the metric.

The metric (\ref{6}) describes a flat 5D manifold ($R^A_{BCD}=0$) in apparent vacuum ($G_{AB}=0$). We consider a diagonal metric 
because we are dealing only with gravitational effects, which are the important ones in the global evolution for the universe. To 
describe neutral matter in a 5D geometrical vacuum (\ref{6}) we can consider the Lagrangian
\begin{equation}\label{1}
^{(5)}{\rm L}(\varphi,\varphi_{,A})= -\sqrt{\left|\frac{^{(5)}
g}{^{(5)}g_0}\right|} \  ^{(5)}{\cal L}(\varphi,\varphi_{,A}),
\end{equation}
where $|^{(5)}g|=\psi^8 e^{6N}$
is the absolute value of the determinant for the 5D metric tensor with
components $g_{AB}$ given by (\ref{6})  and $|^{(5)}g_0|=\psi^8_0 e^{6N_0}$ is a constant of dimensionalization determined by 
$|^{(5)}g|$ evaluated at $\psi=\psi_0$ and $N=N_0$.
In this work we consider $N_0=0$, so that
$^{(5)}g_0=\psi^8_0$. Here, the index $`` 0 "$ denotes the values
at the end of inflation (i.e.,
when the scale factor of the universe satisfies $\ddot b =0$). Furthermore, we consider an action
\begin{equation}\label{action}
I=-\int d^{4}xd\psi\,\sqrt{\left|\frac{^{(5)}
g}{^{(5)}g_0}\right|} \ \left[\frac{^{(5)}R}{16\pi G}+ ^{(5)}{\cal L}(\varphi,\varphi_{,A})\right],
\end{equation}
for a scalar field $\varphi$, which is minimally coupled to gravity. Here, $^{(5)}R$ is the 5D Ricci scalar, which of course, is 
zero for the 5D flat metric (\ref{6}) and $G$ is the gravitational constant,
which we consider independent of the extra dimension.
A different treatment was considered, for example, in \cite{bu}.
Since the 5D metric (\ref{6}) describes a manifold in apparent vacuum, the Lagrangian density ${\cal L}$ in (\ref{1}) must be only
kinetic in origin,
\begin{equation}\label{1'}
^{(5)}{\cal L}(\varphi,\varphi_{,A})
= \frac{1}{2} g^{AB} \varphi_{,A} \varphi_{,B}.
\end{equation}
Considering the metric (\ref{6}) and the Lagrangian (\ref{1}), we obtain the equation of motion for $\varphi$,
\begin{equation}\label{df}
\left(2\psi \frac{\partial\psi}{\partial N}+ 3 \psi^2 \right)
\frac{\partial\varphi}{\partial N}
+\psi^2 \frac{\partial^2\varphi}{\partial N^2} - \psi^2 e^{-2N} \nabla^2_r\varphi
-4\psi^3 \frac{\partial\varphi}{\partial\psi} - 3\psi^4 \frac{\partial N}{
\partial\psi} \frac{\partial\varphi}{\partial\psi} - \psi^4 \frac{\partial^2
\varphi}{\partial\psi^2} =0,
\end{equation}
where $\partial N / \partial\psi$ and $\partial\psi / \partial N$ are zero because the coordinates $(N,\vec{r},\psi)$ are 
independent.\\

\subsection{5D normalization}

Equation (\ref{df}) can be written as
\begin{equation}
\stackrel{\star \star}{\varphi}+3\stackrel{\star}{\varphi}-e^{-2 N}\nabla_{r}^{2}\varphi -\left[4\psi 
\frac{\partial\varphi}{\partial\psi}+\psi^{2} \frac{\partial^{2}\varphi}{\partial\psi^{2}}\right]=0,
\label{ec2}
\end{equation}
where the over-star denotes derivative with respect to $N$ and $\varphi \equiv \varphi(N,\vec{r},\psi)$. Now we proceed with the 
quantization procedure for the scalar field $\varphi (N,\vec{r},\psi)$ in the standard way. We start establishing the commutator 
relation between
$\varphi$ and $\Pi^N = \partial {\cal L} / \partial
\varphi_{,N} = g^{NN} \varphi_{,N}$ that results
\begin{equation}
\left[\varphi(N,\vec r,\psi), \Pi^N(N,\vec{r'},\psi')\right] =ig^{NN} 
\delta^{(3)}(\vec r-
\vec{r'}) \delta(\psi - \psi'),
\end{equation}
being $g^{NN} = \psi^{-2}$.
In order to simplify the structure of (\ref{ec2}), we find convenient to express $\varphi$ as $\varphi =\chi e^{-3N/2}\left(\psi_{0}/ 
\psi \right)^{2}$. Thus, equation (\ref{ec2}) yields 
\begin{equation}
\stackrel{\star \star}{\chi}-\left[e^{-2 N}\nabla_{r}^{2} +\left(\psi^{2} \frac{\partial^2}{\partial\psi^2} +\frac{1}{4}\right) 
\right]\chi =0,
\label{ec3}
\end{equation}
which is a 5D generalized Klein-Gordon like equation for the field $\chi (N,\vec{r},\psi)$.
The field $\chi (N,\vec{r},\psi)$  can be  Fourier expanded as
\begin{eqnarray} \label{ec4}
\chi (N,\vec{r},\psi)&=& \frac{1}{(2\pi)^{3/2}}\int d^{3} k_{r} \int d k_{\psi} \left[a_{k_{r}k_{\psi}} e^{i(\vec{k_r} \cdot 
\vec{r} +{k_\psi}{\psi})}\xi_{k_{r}k_{\psi}}(N,\psi) 
+a_{k_{r}k_{\psi}}^{\dagger} e^{-i(\vec{k_r} \cdot \vec{r} +{k_\psi} {\psi})}\xi_{k_{r}k_{\psi}}^{*}(N,\psi)\right],
\end{eqnarray}
where the asterisk denotes complex conjugation and $(a_{k_{r}k_{\psi}},a_{k_{r}k_{\psi}}^{\dagger})$ are the annihilation and 
creation operators which satisfy the algebra
\begin{eqnarray}\label{ec5}
\left[a_{k_{r}k_{\psi}},a_{k'_{r}k'_{\psi}}^{\dagger}\right]&=&\delta^{(3)}\left(\vec{k_r}-\vec{k'_r}\right) 
\delta\left(\vec{k_\psi}-\vec{k'_\psi}\right),\\
\label{ec6}
\left[a_{k_{r}k_{\psi}}^{\dagger},a_{k'_{r}k'_{\psi}}^{\dagger}\right]&=&\left[a_{k_{r}k_{\psi}},a_{k'_{r}k'_{\psi}}\right]=0,
\end{eqnarray}
and the following commutation relation between $\chi$ and $\stackrel{\star}{\chi}$ must be fulfilled
\begin{equation} \label{ec7}
\left[\chi \left(N,\vec{r},\psi\right),\stackrel{\star}{\chi}\left(N,\vec{r^{ \prime}}, \psi^\prime\right)\right] 
=i\delta^{(3)}\left(\vec{r}-\vec{r^{\prime}}\right) \delta \left({\psi}-{\psi^{\prime}}\right).
\end{equation}
The commutation relation (\ref{ec7}) can be written in terms of the scalar modes $\xi _{k_{r}k_{\psi}}(N,\psi)$ in the form
\begin{equation} \label{recon}
\xi_{k_{r}k_{\psi}}\left(\stackrel{\star}{\xi}_{k_{r} k_{\psi}}\right)^{*} - \left(\xi_{k_{r} k_{\psi}}\right)^{*} 
\stackrel{\star}{\xi}_{k_{r}k_{ \psi }} = i,
\end{equation}
so that this normalization condition is completely equivalent to (\ref{ec7}). Therefore, a solution for the scalar modes $\xi 
_{k_{r}k_{\psi}}(N,\psi)$ that satisfies (\ref{recon}) corresponds automatically to a solution for $\chi(N,\vec{r},\psi)$ that 
satisfies (\ref{ec7}). By inserting (\ref{ec4}) into (\ref{ec3}) we obtain that the dynamical equation for the modes 
$\xi_{k_{r}k_{\psi}}(N,\psi)$ is given by
\begin{equation} \label{ec8}
\stackrel{\star \star}{\xi}_{k_{r}k_{\psi}} +k_{r}^{2} e^{-2 N} \xi_{k_{r}k_{\psi}} +\psi^{2}\left(k_{\psi}^{2} 
-2ik_{\psi}\frac{\partial}{\partial \psi}-\frac{\partial^{2}}{\partial \psi^{2}} -\frac{1}{4\psi^{2}}\right)\xi_{k_{r}k_{\psi}}=0.
\end{equation}
To solve this equation we propose the ansatz
\begin{equation} \label{ec9}
\xi_{k_{r}k_{\psi}}(N,\psi)=\xi_{k_{r}}^{(1)}(N) \xi_{k_{\psi}}^{(2)}(\psi),
\end{equation}
such that (\ref{ec8}) can be written as two differential equations
\begin{eqnarray} \label{ec10}
\stackrel{\star \star}{\xi}_{k_{r}}^{(1)} +\left[k_{r}^{2} e^{-2 N}-\alpha\right]\xi_{k_{r}}^{(1)}&=&0,\\
\label{ec11}
\frac{d^{2}\xi_{k_{\psi}}^{(2)}}{d \psi^2} +2ik_{\psi}\frac{d\xi_{k_{\psi}}^{(2)}}{d\psi}-\left(k_{\psi}^{2} 
-\frac{(1/4-\alpha)}{\psi^2}\right)\xi_{k_{\psi}}^{(2)}&=&0,
\end{eqnarray}
with $\alpha =2k^2_{\psi} \psi^2$.
Solving (\ref{ec10}) and (\ref{ec11}) we obtain
\begin{eqnarray} \label{ec12}
\xi_{k_{r}}^{(1)}[x(N)]&=&A_{1}{\mathcal H}_{\nu}^{(1)}[x(N)]
+A_{2}{\mathcal H}_{\nu}^{(2)}[x(N)],\\
\label{ec13}
\xi_{k_{\psi}}^{(2)} &=& e^{-i\vec{k_{\psi}} \cdot \vec{\psi}} \left[ B_{1} \psi^{\left(\frac{1}{2} +\sqrt{\alpha}\right)} +B_{2} 
\psi^{\left(\frac{1}{2} -\sqrt{\alpha}\right)}\right],
\end{eqnarray}
where ${\mathcal H}_{\nu}^{(1,2)}[x(N)]={\mathcal J}_{\nu}[x(N)]\pm i{\mathcal Y}_{\nu}[x(N)]$ are the Hankel functions, 
${\mathcal J}_{\nu}[x(N)]$ and ${\mathcal Y}_{\nu}[x(N)]$ are the first and second kind Bessel functions with $\nu =\sqrt{\alpha}$ 
and $x(N)=k_{r} e^{-N}$. Furthermore, the arbitrary constants $A_{1}$, $A_{2}$,  $B_{1}$ and $B_{2}$ are constrained by the 
normalization condition (\ref{recon}) in the following manner
\begin{equation} \label{ec16}
[(A_{1}-A_{2}) (A_{1}+A_{2})]\left[B_{1}\psi^{\frac{1}{2} +\nu} +B_{2} \psi^{\frac{1}{2} -\nu}\right]^{2} = \frac{\pi}{4}.
\end{equation}
Choosing the generalized Bunch-Davies vacuum, given in this case by $A_{1}=0$ and $B_{1}=0$ in the above expression, we find that 
the only way to obtain a constant $A_2$ is setting
$\nu=\sqrt{\alpha} =1/2$.
This gives $A_{2}=i\sqrt{\pi}/ (2B_{2})$.
Hence, the solution of (\ref{ec8}) with the normalized condition
(\ref{ec16}), which ensures the quantization of $\varphi$,
is given by
\begin{equation}
\xi_{k_{r}k_{\psi}}(N,\psi)=e^{-i{k}_{\psi} {\psi}} \  \tilde\xi_{k_r}(N),
\end{equation}
being $\tilde\xi_{k_r}(N) i(\sqrt{\pi}/2) {\mathcal H}_{1/2}^{(2)}\left[k_{r}e^{-N}\right]$.
This result possesses great importance and deserves a careful interpretation. The condition (\ref{ec16}),
which guarantees the normalization (and hence the quantization) of the field
$\varphi$, requires that $\alpha = 2 k^2_{\psi} \psi^2$ to be a constant.
This fact can be interpreted in the following manner: the field $\chi $ varies along
the coordinate $\psi$ by changing its momentum in that direction in such
a way that
the product $k_{\psi} \psi$ remains constant (this means that $k_{\psi}$ decreases
as $\psi$ increases or viceversa). A consequence of this is that
the coordinate $\psi $ results to be irrelevant for the
Fourier expansion of $\chi $, but is relevant for $\varphi =\left(\psi _0 / \psi\right)^2 e^{-3 N/2} \chi(N, \vec r)$.

Therefore, the field $\chi$ in equation (\ref{ec4}) now simplifies to
\begin{equation}\label{ecu10}
\chi(N,\vec r, \psi) \chi(N,\vec r) = \frac{1}{(2\pi)^{3/2}} {\Large\int}
d^3k_r {\Large\int} dk_{\psi} \left[
a_{k_r k_{\psi}} e^{i \vec{k_r}.\vec r} \tilde\xi_{k_r}(N) + c.c.\right],
\end{equation}
and thus the field $\varphi$ is given by
\begin{equation}                    \label{24}
\varphi(N,\vec r, \psi) = e^{-3N/2} \left(\frac{\psi_0}{\psi}\right)^2
\chi(N,\vec r),
\end{equation}
with $\chi(N,\vec r)$ given by equation (\ref{ecu10}).
Note that exponentials $e^{\pm i k_{\psi} \psi}$ disappear in
$\chi(N,\vec r)$, which in turn implies that this field is independent of $\psi$. This is very important
because $\varphi(N, \vec r,\psi)$, and hence gravity, propagates
on the 3D spatially isotropic space $r(x,y,z)$, but not on the additional
space-like coordinate $\psi$. This should explain why
matter only propagates on the 4D submanifold described by the coordinates
$(N,\vec r)$.
Furthermore, expression (\ref{24}) is a consequence of the
normalization in (\ref{ec16}), which gives that $\partial\varphi /
\partial\psi = -2 \psi^{-1} \varphi$, because $\alpha $
is a constant of the spectrum such that $k^2_{\psi} = \psi^2/8$.
This fact should be very important in the equation of motion for gravitons
(described by some equation similar to (\ref{ec8})), because
it implies a fixed mass and eliminates the exchange of extra light
particles in the gravitational interactions. However, this is not
the subject of this paper, but the reader can see a different mechanism
to eliminate these light states in Ref.~\cite{bu}.

Throughout the following sections, we study the implications
of this 5D dynamics as seen by 4D observers in a power-law expanding
universe.

\subsection{Effective 4D Power law expansion}

In what follows we extend the formalism, recently introduced by M. Bellini \cite{ref1},
by studying the general case of a variable Hubble parameter.
In this model, developed from a 5D vacuum, the expansion
of the universe is governed by a scalar field, which has been
sliding down its potential energy hill along the entire the evolution
of the universe. The relevance of this model is that the effective
4D dynamics of the field $\varphi $ only depends of the Hubble
parameter $h$, which is a cosmological  observable. Furthermore,
the effective 4D potential $V(\varphi )$ has a geometrical origin.
In particular, in this paper we study
the inflationary expansion and
the radiation dominated epoch. The later, rather than inflation,
describes a decelerated expansion of the universe.
We start by considering the metric (\ref{6}) with the coordinates and
frames transformations
\begin{equation}         \label{trans}
t = {\Large\int} \psi(N) dN,
\qquad R=r\psi, \qquad L=\psi(N) e^{-\int dN/u(N)},
\end{equation}
\begin{eqnarray}
&& U^N = \frac{u(N)}{\psi \sqrt{u^2(N) -1}} \rightarrow
\hat{U}^t = \frac{2u(t)}{\sqrt{u^2(t) -1}}, \\
&& U^r =0 \rightarrow \hat{U}^R = -\frac{2 R h}{\sqrt{u^2(t)-1}}, \\
&& U^{\psi}= -\frac{1}{\sqrt{u^2(N) -1}} \rightarrow \hat{U}^{L} 0,
\end{eqnarray}
such that, for the dynamical foliation $\psi(t)=1/h(t)$ and $u(t) = 1/(1+q(t))$ (being $h(t)$
the effective Hubble parameter), we obtain an effective 4D
Friedmann-Robertson-Walker metric
\begin{equation}\label{frw}
dS^2 \rightarrow ds^2 = \epsilon
\left(dt^2 - e^{2\int h(t) dt} dR^2\right).
\end{equation}
Here $q(t)=-\ddot{b}b/\dot{b}^{2}$ is the deceleration parameter, the overdot denotes  derivative with respect to the time and 
$b(t)$ is the scale factor of the universe \cite{madbe}. The following geodesic dynamics is fulfilled in both reference 
systems\footnote{For a more complete description see \cite{MB}.}
\begin{eqnarray}
&& \frac{dU^C}{dS} = -\Gamma^C_{AB} U^A U^B, \qquad g_{AB} U^A U^B=1, \\
&& \frac{d\hat{U}^C}{dS} = -\hat{\Gamma}^C_{AB} \hat{U}^A \hat{U}^B,
\qquad \hat{g}_{AB} \hat{U}^A \hat{U}^B=1.
\end{eqnarray}
The metric (\ref{frw}) has an effective 4D nonzero scalar curvature
$^{(4)}{\cal R} = 6(\dot h + 2
h^2)$ and a metric tensor with components $g_{\mu\nu}$. The absolute value of the determinant for this tensor is 
$|^{(4)}g|=(b/b_0)^6$.
The Lagrangian density in this new frame was obtained in a previous
work \cite{MB} and it has the form
\begin{equation}
^{(4)} {\cal L}\left[\varphi(\vec{R},t), \varphi_{,\mu}(\vec{R},t)\right]
= \frac{1}{2} g^{\mu\nu} \varphi_{,\mu} \varphi_{,\nu}
- V(\varphi), \label{aa}
\end{equation}
where
\begin{equation}
V(\varphi)=-\left. \frac{1}{2} g^{\psi\psi} \varphi_{,\psi}
\varphi_{,\psi}\right|_{\psi=h^{-1}} = \frac{1}{2}\left. \left(\frac{\partial \varphi}{
\partial \psi}\right)^2 \right|_{\psi = h^{-1}} 2 h^2(t) \varphi^2(t,\vec R, L)
\end{equation}
is the effective 4D induced potential for the scalar field $\varphi(t,\vec{R},L)$.

On the other hand, the 4D expectation
values for energy density $\rho$ and the pressure ${\rm p}$ are given by \cite{PLB}
\begin{eqnarray}
&& 8 \pi G \left<\rho\right> = 3 h^2, \label{ei1}   \\
&& 8\pi G \left<{\rm p}\right> = -(3h^2 + 2 \dot h). \label{ei2}
\end{eqnarray}
where $G=M_{p}^{-2}$ is the gravitational constant and $M_{p}= 1.2\times 10^{19}\,GeV$ is the Planckian mass.
Furthermore, the equation of state that describes the universe is $\left<{\rm p}\right>=-\left[1-2/ 
(3u(t))\right]\left<\rho\right>$,
and from the hyperbolic condition $g_{AB} U^A U^B=1$ we obtain
that the 4D evolution of the universe in terms of $b(t)$ is described by
\begin{equation}
1+\frac{3}{\left(1-\frac{\ddot b b}{\dot b^2}\right)^2} =4 r^2 \left(\frac{
b}{b_0}\right)^2.
\end{equation}
This equation makes an analytical treatment a hard nut to crack. Thus, we restrict ourselves to the case where
$u(t)$ is approximately a constant: $u(t)\simeq p$, where
$p$ describes the power of the expansion for the scale factor $b \sim t^p$.

During inflation, the universe is accelerated ($q<0$) and $u(t) \gg 1$, thus
$\epsilon=1$. However, in epochs of decelerated expansion, $0 < u(t) <1$
and $\epsilon =-1$ in the metric (\ref{6}).
The change of signature of $\epsilon$ means that $dS^2$ changes
sign. This possibility was first considered
by Davidson and Owen \cite{DO},
who studied elementary particles as higher dimensional
tachyons. In the model studied here, causality is affected
during the inflationary epoch, but not when the expansion
of the universe is decelerated.
The possibility that
causality is violated during inflation has been
studied by many other authors as well \cite{Sami}.
In our case we have in mind an early accelerated universe (inflationary expansion),
which at later times is decelerated.
During inflation $dS^2<0$ so that causality is affected, but after
inflation ceases (i.e., when the universe is decelerated), causality is restored,
$dS^2>0$.

The expectation value for the energy density is
\begin{equation}\label{ga1}
\left< \rho\right> = \left< \frac{\dot{\varphi}^2}{2}+\frac{b_{0}^2}{2 b^2}\left(\vec{\nabla}\varphi\right)^{2} + 
V(\varphi)\right>,
\end{equation}
where the brackets denote the 4D expectation value.\\
Considering  equation (\ref{ec2}) and the transformation
of coordinates (\ref{trans}), the effective 4D equation of motion for $\varphi$ is
\begin{equation} \label{ga2}
\ddot{\varphi} + \left(3h-\frac{\dot h}{h}\right)
\dot{\varphi} - e^{-2\int h(t)\,dt}\nabla_{R}^{2}\varphi - \left. \left[\frac{4}{\psi} \frac{\partial\varphi}{\partial\psi}
+ \frac{\partial^2\varphi}{\partial\psi^2}\right]
\right|_{\psi=h^{-1}} = 0,
\end{equation}
and the effective 4D value for $V'(\varphi)$ is
\begin{equation}\label{ga3}
V'(\varphi)\Bigg|_{4D}=-\left[\frac{\dot{h}}{h}\frac{\partial}{\partial t} - 2h^{2}\right]\varphi.
\end{equation}
Now we apply the transformation
\begin{equation} \label{ga4}
\varphi(t,\vec{R}) = \chi(t,\vec{R})e^{-\frac{1}{2}\int\left(3h - \frac{\dot{h}}{h}\right)\,dt} = 
\chi(t,\vec{R})e^{-\frac{3}{2}\int h\,dt}\left(\frac{h}{h_0}\right)^{1/2},
\end{equation}
in order for equation (\ref{ga2}) to become a 4D Klein-Gordon equation
for $\chi$:
\begin{equation} \label{ga5}
\ddot{\chi} - \left[e^{-2\int_{t_0}^{t} h(t')\,dt'}\nabla_{R}^{2} + \frac{h^2}{4} + \frac{3}{4}\left(\frac{\dot{h}}{h}\right)^{2} 
- \frac{1}{2}\left(\frac{\ddot{h}}{h}\right)\right]\chi = 0.
\end{equation}
Here $t_0$ is the time at end of inflation.
On the other hand we have
\begin{equation}\label{ga6}
\chi(t,\vec{R},L) \equiv
\chi(t,\vec{R})\frac{1}{(2\pi)^{3/2}}\int d^{3}k_{R}\int dk_{\psi}\left[a_{k_{R}}e^{i\vec{k_{R}}\cdot\vec{R}}
\tilde\xi_{k_{R}}(t) + c.c.\right]\delta\left(\vec{k_{\psi}}-\vec{k_{L}}\right),
\end{equation}
where $L=\psi_{0}=cte$. Hence the equation of motion for the time dependent modes $\tilde\xi_{k_{R}}(t)$ is
\begin{equation}\label{ga7}
\ddot{\tilde\xi}_{k_{R}} + \left[k_{R}^{2}e^{-2\int h\,dt} - \frac{h^2}{4} - \frac{3}{4}\left(\frac{\dot{h}}{h}\right)^{2} + 
\frac{1}{2}\left(\frac{\ddot{h}}{h}\right)\right]\tilde\xi_{k_{R}} = 0.
\end{equation}
This means that the model excludes non-expanding cosmological models. Besides, all terms inside the brackets have a geometrical 
origin because they are induced by the fifth coordinate.\\

\section{Some examples}

To illustrate the formalism, we study some examples of power-law expanding
universes which are relevant for cosmology.

\subsection{Power-law inflation}

We first consider a power law inflationary
expansion for the universe
\begin{equation}\label{ga8}
h(t)=\frac{p}{t},\qquad p>1.
\end{equation}
Equation (\ref{ga7}) for the $k_R$-modes in power-law inflation is
\begin{equation}\label{ga9}
\ddot{\tilde\xi}_{k_R} +
\left[\frac{\alpha^{2}_0t^{2(1-p)}-\beta^2}{t^2}\right] \tilde\xi_{k_R}=0,
\end{equation}
with $\alpha_0 = k_{R} t_{0}^{p}$, $\beta^2 = (p^{2}-1)/4$.
The general solution of (\ref{ga9}) is
\begin{equation}\label{ga10}
\tilde\xi_{k_R}[x(t)] = \sqrt{t}\left(C_{1}{\mathcal H}_{\nu}^{(1)}[x(t)]+C_{2}{\mathcal H}_{\nu}^{(2)}[x(t)]\right),
\end{equation}
being $\nu=p/[2(p-1)]$ and $x(t)=(\alpha_0 t^{1-p})/(p-1)$. Hence, when the normalization condition for $\tilde\xi_{k_R}(t)$ and 
the Bunch-Davies vacuum are both considered, i.e., $C_{1}=0$, $C_{2}=i\sqrt{\pi/[4(p-1)]}$, we obtain
\begin{equation}\label{ga11}
\tilde\xi_{k_R}(t) = i\sqrt{\frac{\pi}{4(p-1)}}\,\sqrt{t}\,
{\mathcal H}_{\nu}^{(2)}\left[\frac{\alpha_0 t^{1-p}}{p-1}\right],
\end{equation}
where the condition $\nu=1/2$ is fulfilled only for
$p \rightarrow \infty$ in inflationary models.
Note that this result is very similar to the one obtained in the power law standard inflationary approach.\\

Once the time dependent 4D-modes $\tilde\xi_{k_R}(t)$ are known, we can calculate the effective 4D super Hubble squared 
fluctuations
$\left< \varphi^2 \right>$.
For this purpose, we must remember the small argument
limit for the second kind Hankel function: ${\mathcal H}_{\nu}^{(2)}[x]\simeq (x/ 2)^{\nu}/ \Gamma (1+\nu)- (i/ \pi)\Gamma 
(\nu)\left(x/ 2\right)^{-\nu}$.
Note that in this case $\nu < 0$ because $p > 1$, so we can use ${\mathcal H}_{\nu}^{(2)}[x]\simeq (x/ 2)^{\nu}/ \Gamma (1+\nu)$ 
and hence $\left< \varphi^{2}(\vec{R},t)\right>$ can
be calculated on the infrared (IR) sector from the expression
\begin{equation}\label{12}
\left<\varphi^{2}(t,\vec{R})\right>_{IR}
= \frac{2^{\frac{p}{p-1}} (p-1)^{\frac{1}{p-1}}}{
8 \pi \Gamma^2\left[\frac{p-2}{2(p-1)}\right]}
t^{2p}_0 t^{-2p}
\int_{0}^{\theta k_H}\frac{dk_{R}}{k_R}\,k_{R}^{\frac{2p-3}{(p-1)}},
\end{equation}
where $\theta =k^{(IR)}_{max} /k_p \ll 1$ is a dimensionless constant
(of order $\sim10^{-4}-10^{-3}$),
$k^{(IR)}_{max} = k_H(t_*)=\left.e^{\int h dt} h\right|_{t=t_*}p \left(t_*/ t_0\right)^p t^{-1}_*$ is the wavenumber
at the moment $t_*$, when the horizon enters, and
$k_p$ is the Planckian wavenumber (i.e., the scale we choose as a cut-off
of all the spectrum).
In other words, $k_H(t_*)$ is the wavenumber
related to the Hubble radius in an expanding universe when the
horizon enters.

\subsection{Radiation dominated universe}

A radiation dominated universe is described by a power $p=1/2$, which
is related with a Hubble parameter $h(t)=1/(2t)$, such that
equation (\ref{ga7}) becomes
\begin{equation}\label{rad}
\ddot{\tilde\xi}_{k_R}
+ \left(\frac{€\alpha^2_1 t + \beta^2_1}{t^2}\right)
\tilde\xi_{k_R} =0,
\end{equation}
with $\alpha_1 = k_R t^{1/2}_0$ and $\beta^2_1=3/ 16$. The
general solution for this equation is
\begin{equation}
\tilde\xi_{k_R}(t) = G_1 \  \sqrt{t}
{\cal H}^{(1)}_{\nu_1}[2\alpha_1 t^{1/2}]
+ G_2 \  \sqrt{t} {\cal H}^{(2)}_{\nu_1}[2\alpha_1 t^{1/2}],
\end{equation}
being ($G_1$, $G_2$) constants and $\nu_1 = \sqrt{1-4\beta^2_1}$.
Hence, when the normalization condition for $\xi_{k_R}(t)$ and
the Bunch-Davies vacuum are considered, we obtain
$G_1=0$ and $G_2=\sqrt{\pi/ 2}$. Thus, the normalized solution for
equation (\ref{rad}) is
\begin{equation}
\tilde\xi_{k_R}(t)
= \sqrt{\frac{\pi}{2}} \sqrt{t} {\cal H}^{(2)}_{\nu_1}
[2\alpha_1 t^{1/2}],
\end{equation}
where in our case $\nu_1 = 1/2$. In this fashion, on
super Hubble scales we obtain
\begin{equation}\label{fluc1}
\left< \varphi^2\right>_{IR}
= \frac{\Gamma^2[1/2]}{4 \pi^3} \left(\frac{t_0}{t}
\right) {\Large\int}^{\theta k_H}_0 \frac{dk_R}{k_R} \  k^2_R,
\end{equation}
which, after integration yields
\begin{equation}\label{ult}
\left< \varphi^2\right>_{IR} = \frac{\theta^2 \Gamma^2[1/2]}{32 \pi^3}
t^{-2}.
\end{equation}
Note the difference between the solutions of equations (\ref{ga9}) and (\ref{rad}).
Solutions of (\ref{ga9})
are unstable outside the Hubble horizon and stable
inside it. However, solutions of (\ref{rad}) are stable over the entire
the spectrum. Thus, inflationary dynamics is necessary to explain
why the universe is larger than the Hubble horizon.

\section{Final Comments}

In this work we have extended a previously introduced
nonperturbative scalar field formalism for a noncompact KK theory
to a power-law expanding universe. In particular, we have examined
some relevant cases in cosmology; inflationary dynamics ($p>1$)
and radiation dominated (with $p=1/2$) universes, respectively. We
have found that the $\left<\varphi^2\right>_{IR}$ spectrum is
scale invariant only during inflation for $p=3/2$. This formalism
has the advantage, with respect to the semi-classical approach
studied in \cite{MB}, that it provides an exact (nonperturbative)
treatment for the dynamics of the inflaton field (the scalar
field). Furthermore, it provides a consistent treatment for the
effective 4D dynamics of the universe in models governed by a
single scalar field. Note that the dynamics is induced
geometrically through the potential in the frame $\hat U^t = 2u /(
\sqrt{u^2-1})$, $\hat U^R = - 2 Rh / (\sqrt{u^2-1})$, $\hat U^L
=0$, where the fifth coordinate $L=\psi_0$ remains constant. This
constant is given by the Hubble horizon at the end of inflation.
The potential is quadratic in $\varphi$ but its squared mass
depends on the rate of expansion of the universe (the Hubble
parameter). In epochs where the expansion is accelerated the
signature of the squared mass is positive, but when the expansion
is decelerated it shifts to a negative signature.

Since tensor metric fluctuations $h_{AB}$ are not the subject of this paper, in our analysis we excluded possible short-range 
modifications of Newton's gravitational law due to the 5D graviton propagator.
For a discussion about astrophysical implications of the existence
of extra dimensions see \cite{u2,u3}.

Finally, the theory can explain why inflationary dynamics is necessary
to explain the existence of an universe larger than the Hubble horizon,
because the modes related to cosmological scales of the
quantum field $\chi$ grows superluminally for a scale factor
$b \sim t^p$, with $p>1$. However, for a radiation dominated
universe the modes are stable over the entire the spectrum. This
is a remarkable result of this paper: \emph{during the radiation dominated
epoch the squared $\varphi$-fluctuations are stable over all scales, rather
in models where the universe is (as inflation) accelerated.}
It is evident the difference between the result (\ref{ult}) and
those obtained in \cite{ulti}, where
$\left< \varphi^2\right>_{IR} \sim t^{-3/2}$. However, as
was demonstrated in \cite{ulti} the stochastic
approach is only valid for $p\gg 2$.

\section*{Acknowledgements}
\noindent
AR acknowledges CIC-UMSNH (Mexico) for financial support under project 4.22. JEMA acknowledges CNPq-CLAF and UFPB (Brazil) for financial support. MB acknowledges CONICET and UNMdP (Argentina) for financial support.

\end{document}